\documentclass[a4paper,12pt]{article} 
\usepackage{amsmath}  
\usepackage{amsfonts} 
\usepackage{graphicx} 
\usepackage{latexsym}

\begin{document}

\title{The bizarre anti--de Sitter spacetime}

\author{Leszek M. Soko\l{}owski,\\
Astronomical Observatory, Jagiellonian University,\\
 Orla 171, Krak\'ow 30-244, Poland,\\
email:lech.sokolowski@uj.edu.pl} 
\date{}
\maketitle

\begin{abstract}
Anti--de Sitter spacetime is important in general relativity and modern field theory. 
We review its geometrical features and properties of light signals and free particles 
moving in it. Applying only elementary tools of tensor calculus we derive \textit{ab 
initio\/} all these properties and show that they are really weird. One finds 
superluminal velocities of light and particles, infinite particle energy necessary to 
escape at infinite distance and spacetime regions inaccessible by a free fall, though 
reachable by an accelerated spaceship. Radial timelike geodesics are identical to 
the circular ones and actually all timelike geodesics are identical to one circle in 
a fictitious five--dimensional space. Employing the latter space one is able to explain 
these bizarre features of anti--de Sitter spacetime; in this sense the spacetime is 
not self--contained. This is not a physical world.\\

\end{abstract}
\maketitle 

\section{Introduction} 
The anti--de Sitter spacetime is one of the simplest and most symmetric 
solutions of Einstein's field equations including the cosmological constant. 
For this reason it is important for general relativity and it has its own 
mathematical relevance. After 1998 this spacetime has drawn attention of high 
energy physicists due to the conjectured anti--de Sitter space/conformal field 
theory (AdS/CFT) correspondence suggesting that fundamental particle interactions 
may be described in geometrical terms with the aid of this spacetime \cite{Hu}. This idea 
has given rise to a great number of works on this spacetime which take into account 
only those geometrical features of it that are relevant in this quantum field theory 
aspect and seem to disregard all its other properties. We shall not discuss the 
correspondence, we wish only to emphasize that this spacetime, which has become one 
of the most fundamental spacetimes in physics, has rather bizarre geometrical properties 
and is weird also from the physical viewpoint. By the latter we mean motions of 
material (classical) bodies and propagation of light signals in this background.\\

In this spacetime almost everything is bizarre including its name. In the older literature,
particularly mathematical, it was termed \textit{de Sitter spacetime of the second kind\/} 
and the current name has been given to it to stress that its geometrical properties are 
opposite to those of de Sitter spacetime (which was studied earlier and more frequently 
as it better fits our intuition) though at first sight the two spaces should be similar. 
(To the best of our knowledge the term appeared for the first time in Ref. \cite{CM}).  
These bizarre properties were discovered by mathematicians rather long ago and exist in 
the literature which is now not easy to find. This is why this paper is written: its 
purpose is to collect and present in a possibly systematic way those features of the 
spacetime which are geometrically and physically important and can be expounded in 
almost elementary terms without resorting to sophisticated mathematics. In consequence 
its contents are hardly new, nonetheless 
we give rather few references. We find it easier to explicitly derive \textit{ab 
initio\/} each result than to seek it in the dispersed literature; thus in most cases 
we cannot pretend to originality. Some very recently published and unpublished results are 
presented in sections 7, 8, 9 and in Appendix.\\
We first give the geometrical construction of the spacetime and show some of its global 
features. Then we focus our interest on motions: what an observer would see if he 
occurred to be there. We present all these effects in analytic form and our figures 
are simple diagrams illustrating these expressions. The reader interested in various 
images of the spacetime is referred to Ref. \cite{GP}. We assume that the reader is 
 familiar with fundamentals of general relativity and tensor calculus.

\section{Geometrical construction and various coordinate systems}
The name of the spacetime will be abbreviated to \textit{AdS space\/} and the term 
\textit{space\/} will mean \textit{spacetime\/} whenever there will be no risk of 
confusing it with the \textit{physical space\/} of the spacetime. AdS space may be 
defined in any number of spacetime dimensions equal to or larger than 2. Here we will 
be dealing only with the physical case of 4 dimensions. First one introduces an 
auxiliary unphysical 5--dimensional flat space $\mathbf{R}^{3,2}$ with Cartesian 
coordinates $(U,V,X,Y,Z)$ having two timelike dimensions $U$ and $V$ and three 
spatial ones $X$, $Y$, $Z$. Accordingly, the line element (the square of the 
spacetime interval) or the metric is 
\begin{equation}\label{n1}
ds^2=dU^2+dV^2-dX^2-dY^2-dZ^2.
\end{equation}
AdS is defined as a 4--dimensional hypersurface in $\mathbf{R}^{3,2}$ given by the 
equation 
\begin{equation}\label{n2}
U^2+V^2-X^2-Y^2-Z^2=a^2.
\end{equation}
The constant $a$ has dimension of length and determines, as we shall see, the curvature 
scale of AdS. The hypersurface is the locus of points equidistant (in this metric) to 
the origin of the Cartesian coordinate system and is legitimately termed 
\textit{pseudosphere\/}. Yet if one takes the equation $U^2+V^2-X^2=a^2$ in the 
euclidean 3--space $(U,V,X)$, the equation represents a one--sheeted hyperboloid and 
by this analogy the hypersurface of eq. (2) is also dubbed \textit{hyperboloid\/}. One 
can parametrize points of the pseudosphere by means of four parameters which are so 
chosen that eq. (2) holds identically. Different parameterizations correspond to 
distinct coordinate systems on AdS. Here we present 5 different systems and each of 
them is most suitable for displaying a distinct geometrical feature.\\
Before doing it a comment on a distinction between reference frames and coordinate 
systems is in order. A reference frame is an ordered structure of material bodies, 
either point particles or extended bodies (rigid or not), covering the entire space of 
the spacetime, together with an infinite set of clocks densely located in the space and 
remaining at rest with respect to nearby bodies of the frame (in general the clocks and 
the bodies to which they are attached may move with respect to distant bodies of the frame---
in the sense that the distance between them may vary in time)\footnote{This definition is 
intuitive, a precise one is more complicated.}. The reference frame is a physical system which, 
at least in principle, can be built out of massive particles and which is the essential 
structure to make any physical measurements and to label spacetime points (events). The 
fundamental example is any inertial frame of reference in special relativity, being a 
dense infinite grid of rigid rods, equipped with clocks located at the intersection points 
of the grid; the whole system is free of accelerations and nonrotating. In a curved 
spacetime the collection of reference frames must be much wider and clearly there are no 
inertial frames. Yet a coordinate system is a purely mathematical way of labelling points 
in the spacetime (in the mathematical language it is a coordinate chart on a differential 
manifold, with all the charts forming the atlas). Each physical reference frame allows to 
introduce infinite number of coordinate systems. For instance, in an inertial frame, the 
standard Cartesian coordinates $(t,x,y,z)$, where $t$ is the physical (i.~e.~proper) 
time measured by clocks in this frame, one can introduce coordinates $(t',r,\theta, \phi)$, 
where $(r,\theta,\phi)$ are curvilinear spatial coordinates defined as given functions of 
$x$, $y$, $z$, e.~g.~the spherical ones and $t'=f(t)$ with monotonously growing function $f$. 
We emphasize that to assign coordinates to points in a physical spacetime one must apply a 
material reference frame and in this sense most of coordinate systems that are used are 
connected to some frame. However the freedom to mathematically construct coordinate systems 
is larger than it is allowed by reference frames. This means that there are coordinate systems 
which are not generated by a reference frame, e.~g.~null coordinates defined in terms of a 
,,null frame''; these coordinates are useful in some calculations, but they are not 
measurable.\\

1. Parameters $(t,r,\theta,\phi)$. Points of AdS in $\mathbf{R}^{3,2}$ are 
represented by
\begin{eqnarray}\label{n3}
U & = & a\sin \frac{t}{a}\cosh\frac{r}{a}, \quad V=a\cos\frac{t}{a}\cosh\frac{r}{a},
\nonumber\\
X & = & a\sinh\frac{r}{a}\sin\theta\cos\phi, \, Y=a\sinh\frac{r}{a}\sin\theta\sin\phi, 
\, Z=a\sinh\frac{r}{a} \cos\theta,
\end{eqnarray}
here $-\pi a<t<\pi a$, $r\geq 0$ and $0\leq \theta\leq \pi$ and $0\leq \phi<2\pi$ are 
ordinary angular coordinates on the 2--sphere $S^2$. Inserting eq. (3) into eq. (2) one 
finds that it holds identically. Yet inserting eq. (3) into the line element (1) one finds 
that the square of the interval between two close points, $(t,r,\theta, \phi)$ and 
$(t+dt, r+dr,\theta+d\theta,\phi+d\phi)$ on the pseudosphere is
\begin{equation}\label{n4}
ds^2=\cosh^2\frac{r}{a}\,dt^2-dr^2-a^2\sinh^2\frac{r}{a}\,(d\theta^2+\sin^2\theta\,
d\phi^2).
\end{equation}
By comparison with the line element in Minkowski space in spherical coordinates $(t,r,
\theta,\phi)$ one identifies $t$ as a time coordinate and $r$, $\theta$, $\phi$ as 
spatial coordinates and the angles $\theta$ and $\phi$ determine the metric on the 
unit sphere $S^2$ as $dl^2=d\theta^2+\sin^2\theta\,d\phi^2\equiv d\Omega^2$. Then $r$ 
is interpreted as a \textit{radial coordinate\/}, but this term does not determine 
the coordinate uniquely. The radial coordinate in the euclidean 3--space $E^3$ has 
two features: if points of a sphere have the radial coordinate $r=r_0$, then i) the 
length of the equator is $2\pi r_0$ (and the area of the sphere is $4\pi r_0^2$) 
and ii) the radius of the sphere, i.~e.~the distance of each of its points to the 
centre is $r_0$. These two features cannot hold together in a curved space and one 
must choose between them. A space is spherically symmetric (only then the notion 
of the radial coordinate makes sense) if there exist coordinates, frequently denoted 
$(t,r,\theta,\phi)$, such that $\theta$ and $\phi$ are the angular coordinates on the 
sphere and the full metric depends on the angles via only one term $g_{22}(t,r)
\,d\Omega^2$ (actually the correct mathematical definition is more sophisticated 
and we omit it); then $r$ deserves the name ,,radial''. Any transformation 
$r'=f(r)$ with $df/dr\neq 0$ gives rise to another radial variable. In the 
metric eq. (4) the coordinate $r$ is equal to the radius of the sphere, whereas 
the length of the equator is $2\pi\sinh r/a$. The following two coordinate systems 
differ from that of eq. (4) only by the choice of the radial coordinate.\\
Notice that the time $t$ has dimension of length  or is measured in ,,light 
seconds''. We do not explicitly introduce the light velocity $c$ here and throughout 
the paper each time coordinate $\tau$ should be interpreted as $c\tau$. \\

2. The transformation $\rho\equiv a\sinh r/a$ yields $\rho\geq 0$ and 
\begin{equation}\label{n5}
ds^2=\frac{\rho^2+a^2}{a^2}\,dt^2-\frac{a^2}{\rho^2+a^2}\,d\rho^2-\rho^2\,d\Omega^2.
\end{equation} 
Here the sphere $\rho=\rho_0$ has the radius equal to the length of the spatial 
curve $dt=d\theta=d\phi=0$, or
\begin{equation}\label{n6}
\int_0^{\rho_0} dl\equiv
\int_0^{\rho_0}\sqrt{-ds^2}=\int_0^{\rho_0} \frac{a\,d\rho}{\sqrt{\rho^2+a^2}}=
a\ln\left(\frac{\rho_0}{a}+\frac{1}{a}\sqrt{\rho_0^2+a^2}\right),
\end{equation} 
whereas the length of the equator is $2\pi \rho_0$. In these coordinates one sees 
that the flat Minkowski space arises in the limit $a\rightarrow\infty$, then 
$\rho$ becomes the ordinary radial coordinate; it is less easy to notice this 
limit in the coordinate $r$ of eq. (4).\\

3. The ,,radial'' angle $\psi$ is defined by $\sinh r/a=\tan\psi$, then 
$0\leq \psi<\pi/2$ and 
\begin{equation}\label{n7}
ds^2=\frac{dt^2}{\cos^2\psi}-\frac{a^2}{\cos^2\psi}\,(d\psi^2 +\sin^2\psi\,
d\Omega^2),
\end{equation} 
now both the radius of the sphere and its circumference do not have their familiar 
forms.\\

The three coordinate systems represent the same physical reference frame and have 
common important features. In the defining equation (2) all the five coordinates range 
from $-\infty$ to $+\infty$ and the transformation (3) preserves this range. This 
implies that the charts (coordinate systems) (4), (5) and (7) cover the entire 
manifold (spacetime) besides the coordinate singularities such as $r=\rho=\psi=0$. 
The hypersurfaces of simultaneity $t=$const form the physical 3--spaces with the 
metric defined as $dl^2\equiv -ds^2$ for $dt=0$. From eq. (4), 
\begin{equation}\label{n8}
dl^2=dr^2+a^2\sinh^2\frac{r}{a}\,d\Omega^2.
\end{equation}
This is Lobatchevsky (hyperbolic) space $H^3$ with coordinates $r$, $\theta$, $\phi$. 
The curvature tensor of $H^3$ is (Greek indices are spacetime ones, $\alpha,\beta,
\mu,\nu=0,1,2,3$ and Latin lower case indices are spatial, $i,j,k=1,2,3$)
\begin{equation}\label{n9}
R^{(3)}_{ijkl}=\frac{R^{(3)}}{6}\,(g_{ik}g_{jl}-g_{il}g_{jk}).
\end{equation}
The curvature scalar $R^{(3)}\equiv g^{ik}g^{jl}R^{(3)}_{ijkl}$ for eq. (8) is equal to 
$R^{(3)}=-6/a^2$ and this property together with eq. (9) is expressed by saying that 
the hyperbolic space $H^3$ is a \textit{space of constant negative curvature\/}. The 
space AdS has an analogous property: its four--dimensional Riemann tensor is given by 
a similar expression,
\begin{equation}\label{n10}
R_{\alpha\beta\mu\nu}=\frac{R}{12}\,(g_{\alpha\mu}g_{\beta\nu}-g_{\alpha\nu}g_{\beta\mu}),
\end{equation}
where its metric $g_{\mu\nu}$ is taken either from eq. (4), (5) or (7) (or any other 
coordinate system) and the 4--dimensional curvature scalar $R=g^{\alpha\mu}
g^{\beta\nu}R_{\alpha\beta\mu\nu}=12/a^2$. Notice that the metric signature is chosen 
here as $(+---)$ since it is more suitable for dealing with timelike worldlines of 
massive particles, whereas in classical field theory the opposite signature is commonly 
used. Altering the signature results in the change of sign of the scalar $R$ and this 
is why AdS space is frequently characterized as a \textit{spacetime of constant negative 
curvature\/}.\\
The metric of eqs. (4), (5) and (7) is time independent, what means that AdS space is 
\textit{stationary\/}. Furthermore, this spacetime is \textit{static\/}, i.~e.~the 
time inversion $t\rightarrow -t$ does not change the form of the metric. The 
gravitational field of a motionless star is static (for instance Schwarzschild field), 
yet a uniformly rotating star generates a stationary field: it is time independent, 
but the time inversion makes the star rotate in the opposite direction and its 
gravitational field is changed (e.~g.~Kerr spacetime).\\
Now we introduce two further coordinate systems describing two different reference 
frames.\\

4. The Poincar\'{e} coordinates $(t',x,y,z)$. Instead of eq. (3) one applies 
\begin{eqnarray}\label{n11}
U & = & \frac{1}{2z}(a^2+x^2+y^2+z^2-t'^2), \quad V=a\frac{t'}{z}, \quad 
X=a\frac{x}{z},
\nonumber\\
Y & = & a\frac{y}{z}, \quad Z=\frac{1}{2z}(a^2-x^2-y^2-z^2+t'^2), 
\end{eqnarray}
then the metric is
\begin{equation}\label{n12}
ds^2=\frac{a^2}{z^2}\,(dt'^2-dx^2-dy^2-dz^2).
\end{equation}
Here $t'$, $x$ and $y$ are real and $z>0$. From eq. (11) one gets $U+Z=a^2/z>0$, 
what implies that these coordinates cover only one half of AdS manifold. The 
other half needs a similar chart with $z<0$. The reference system given in eq. (12) 
is moving with respect to that given in eq. (4) and their coordinate times, $t$ 
and $t'$ are distinct. The expression in the round brackets in eq. (12) represents 
the metric of flat Minkowski space expressed in Cartesian coordinates of an inertial 
reference frame. (At this moment we disregard the derivation of eq. (12) and 
discuss only its final form.) Thus the metric of AdS is proportional to the metric 
of the flat spacetime, the proportionality factor is a scalar function of 
the coordinates. This is a geometrical property of AdS space, valid in all 
coordinate systems. The Poincar\'{e} coordinates are distinguished by making this 
property explicit; it is rather hard to recognize it in other coordinates. By 
,,hard'' we mean that if one uses only the three above mentioned coordinate systems 
(or any other ones) and is unaware that the spacetime is the pseudosphere in 
$\mathbf{R}^{3,2}$ and that it may be parametrized by the Poincar\'{e} coordinates, 
then finding out the transformation to the metric (12) is really difficult. Yet 
showing this property is actually quite easy if one uses the Weyl tensor: this 
tensor is related to the Riemann curvature one and if the proportionality property 
holds for a spacetime, then this tensor (computed in any coordinate system) 
vanishes. In short, if the Weyl tensor is zero, then the metric is proportional to 
the flat one. In this article we shall not apply this tensor. If two 
spacetimes, $M$ and $\bar{M}$, have their metric tensors (expressed in the same 
coordinates) proportional, $\bar{g}_{\mu\nu}(x^{\alpha})=\Omega^2(x)g_{\mu\nu}
(x^{\alpha})$, where $\Omega(x)>0$ is a scalar function, then the two spacetimes are 
\textit{conformally related\/}. Let two conformally related metric tensors be 
introduced on the same spacetime (considered as a ,,bare'' manifold of points), 
then distances between any pair of points expressed in terms of these metrics 
will be different, yet the angles between any two curves are the same in both 
the metrics and this explains why the property is called \textit{conformality\/}. 
AdS is \textit{conformally flat\/}.\\
 The space $t'=\textrm{const}$ in Poincar\'{e} coordinates is conformal to a half 
of euclidean space $E^3$.\\

5. Finally one takes the following parametrization: 
\begin{eqnarray}\label{n13}
U & = & a\cos\frac{\tau}{a}, \quad V=a\sin\frac{\tau}{a}\cosh\chi, \quad 
X=a\sin\frac{\tau}{a}\sinh\chi\sin\theta\cos\phi,
\nonumber\\
Y & = & a\sin\frac{\tau}{a}\sinh\chi\sin\theta\sin\phi, \quad 
Z=a\sin\frac{\tau}{a}\sinh\chi\cos\theta, 
\end{eqnarray}
where $0<\tau<\pi a$ and the radial coordinate $\chi>0$ is dimensionless. The 
metric is now time dependent,
\begin{equation}\label{n14}
ds^2= d\tau^2-a^2\sin^2\frac{\tau}{a}\,(d\chi^2+\sinh^2\chi\,d\Omega^2).
\end{equation}
The coordinates cover only a part of the spacetime since $-a<U<a$ and $V>0$.  
The static nature of AdS becomes now invisible and at first sight these coordinates 
seem to be a mere complication. We shall see below, however, that $(\tau,\chi,\theta,
\phi)$ are \textit{comoving coordinates\/} and reveal an important property of motion 
of free particles. By comparing eqs. (4), (8) and (14) one sees that the space 
$\tau=\textrm{const}$ is the Lobatchevsky space $H^3$.\\
One may introduce a number of other coordinates, but the spherical angles $\theta$ 
and $\phi$ are never altered.

\section{Global properties of the spacetime}
AdS space as the pseudosphere in the ambient $\mathbf{R}^{3,2}$ is unbounded in each 
direction. Yet from eq. (3) one sees that the times $U$ and $V$ are parametrized by a 
\textit{periodic\/} time $t$ on the pseudosphere: the two quadruples, 
$q_1=(t,r,\theta,\phi)$ and $q_2=(t+2\pi a,r,\theta,\phi)$ represent the same point of it. 
 This means that in AdS space, defined as a manifold of points $(t,r,\theta,\phi)$, one 
must identify $q_1$ and $q_2$. More precisely, the range of time is $-\pi a\leq t<\pi a$ 
and points $(-\pi a, r,\theta,\phi)$ and $(+\pi a, r,\theta,\phi)$ are identified. In other 
terms, the coordinate lines of time $t$, where $r,\theta,\phi=\textrm{const}$, are closed 
---they form circles $S^1$. On the other hand the hyperbolic space $H^3$ has topology 
(in the sense of geometrical topology) of euclidean $\mathbf{R}^3$, then the entire AdS 
has the product topology $S^1\times\mathbf{R}^3$. Closed timelike curves are very unpleasant 
from the physical viewpoint. Though it may be argued that they do not break the causality 
and need not to give rise to various paradoxes (,,to kill one's own grandfather''), 
it is desired to remove them if possible. This may be achieved due to the fact that 
the metric (4) (as well as eq. (5) and (7)) is time independent and the periodicity in 
time is invisible. One simply unwraps all time circles $S^1$ and extends them in the 
line of real numbers, now $-\infty<t<+\infty$. Geometrically this means making 
infinite number of turns around the pseudosphere in its time direction. To  avoid 
the periodic identification of points in this direction one discards the 
pseudosphere model and introduces a new spacetime: one discards the whole 
derivation of eq. (4) based on employing the ambient space $\mathbf{R}^{3,2}$ 
and constructing the pseudosphere in it. Instead one defines the manifold as a 
set of points $(t,r,\theta,\phi)$ with $-\infty<t<\infty$ equipped with the 
metric in eq. (4). The coordinate lines of time have now topology $\mathbf{R}^1$ 
and the entire spacetime has topology $\mathbf{R}^4$. This spacetime is called 
a \textit{universal covering space\/} of anti--de Sitter space, in short CAdS. 
In what follows we shall be mainly dealing with CAdS space (unless otherwise is stated).  
It will be quite surprising to see that replacing AdS by CAdS 
space is a merely verbal operation and the latter inherits most of the 
features of the former.\\
In the search for symmetries of CAdS space one may resort to the pseudosphere 
since symmetries are local isometric mappings of the space onto itself preserving 
the form of the metric and do not depend on the topology. Like the ordinary sphere 
in euclidean space, the pseudosphere has as its symmetries the rotational symmetry 
of the ambient space, in this case this is $SO(3,2)$ group, which is analogous to 
$SO(3,1)$ Lorentz group of Minkowski space. This group has 10 parameters, the maximal 
number of symmetries in four dimensions; equally high symmetry is characteristic 
for Minkowski and de Sitter spacetimes. CAdS is \textit{maximally symmetric\/}.\\

An important global property of a spacetime is its structure at infinity. This is termed 
\textit{conformal structure\/} and has been developed in an extended subject presented in 
advanced textbooks \cite{HE, Pe}.                            
Here we need only one, the simplest and most intuitive notion. In Minkowski spacetime the 
boundary of the space $t=\textrm{const}$ for $r\to\infty$ (it is convenient here to use the 
spherical coordinates $(t,r,\theta,\phi)$) is the \textit{sphere at infinity\/}. The 
collection of these spheres for all values of time forms a 3--dimensional hypersurface 
$\mathcal{J}$ being a boundary of the spacetime. To investigate the geometry of 
$\mathcal{J}$ one considers a special metric conformally related to the flat one. The 
resulting geometry of $\mathcal{J}$ is somewhat complicated, whereas the corresponding 
boundary of CAdS space, termed \textit{spatial infinity\/} and also denoted by 
$\mathcal{J}$, is geometrically simpler. The coordinates $(t,\psi,\theta,\phi)$ of eq. 
(7) are most suitable for dealing with the infinity $\psi=\pi/2$. On CAdS space one 
introduces a new metric conformally related to that of eq. (7), $\bar{g}_{\mu\nu}=
\Omega^2g_{\mu\nu}$ with $\Omega=\cos\psi$. In this way one gets a new spacetime with 
the metric
\begin{equation}\label{n15}
d\bar{s}^2= dt^2-a^2\,(d\psi^2+\sin^2\psi\,d\Omega^2).
\end{equation} 
The new spacetime is larger than CAdS space since points $\psi=\pi/2$ are now of its 
regular points, whereas the metric (7) is divergent there. Points $\psi=\pi/2$ of the 
new spacetime form the \textit{conformal spatial infinity\/} $\mathcal{J}$ of CAdS 
space. This hypersurface has the metric (15) with $\psi=\pi/2$, 
\begin{equation}\label{n16}
d\bar{s}^2= dt^2-a^2\,(d\theta^2+\sin^2\theta\,d\phi^2).
\end{equation} 
In a spacetime any hypersurface defined by an equation $f(x^{\alpha})=0$ belongs to one 
of three classes of hypersurfaces depending on the vector $n^{\alpha}$ orthogonal to it: 
if $n^{\alpha}$ is timelike, $n^{\alpha}n_{\alpha}>0$ (according to the signature 
$(+---)$), then the hypersurface is spacelike, if $n^{\alpha}$ is spacelike, 
$n^{\alpha}n_{\alpha}<0$, the hypersurface is timelike (and is a 3--dimensional 
spacetime on its own), finally, if $n^{\alpha}$ is null, $n^{\alpha}n_{\alpha}=0$, it 
lies on the hypersurface to which it is orthogonal and the latter is null. $n^{\alpha}$ 
is the gradient of $f$, $n_{\alpha}=\partial f/\partial x^{\alpha}$. In general the type 
of a hypersurface may change from point to point. In GR we try to avoid this pathological 
behaviour and only consider hypersurfaces which are of the same type everywhere. In the geometry of 
eq. (15) one has $f=\psi-\pi/2$, $n_{\alpha}=(0,1,0,0)$ and $n^{\alpha}n_{\alpha}=
\bar{g}^{\alpha\beta}n^{\alpha}n_{\beta}=\bar{g}^{11}=-1/a^2<0$. The conformal infinity 
$\mathcal{J}$ of CAdS is a \textit{timelike hypersurface\/} and, as is seen from eq. (16), 
it has topology $\mathbf{R}^1\times S^2$, where $S^2$ is the boundary at infinity of the 
space $H^3$. The conformal boundaries of Minkowski and CAdS spaces are different.\\
A null vector in the given metric remains null in all other metrics conformally related 
to that, hence a null line remains null. CAdS space is conformally flat 
(eq. (12)),                
therefore the light cones formed by light rays emitted from any point of that spacetime 
are the same as those in Minkowski space. In particular the straight lines at 
$\pm45$ degrees in the spacetime diagram represent null rays (radially directed photon 
worldlines).\\
Since the infinity $\mathcal{J}$ is actually timelike, the effect is that far future 
cannot be predicted in CAdS space. Suppose one is interested in finding a unique 
solution to Maxwell equations. To this end one chooses a spacelike hypersurface $S$, 
given by $t=t_0$ in some coordinate system, gives the initial data on it (values of the 
electric and magnetic fields at points of $S$) and evolves the data by means of Maxwell 
equations to the future. The value of the electromagnetic field cannot be predicted in 
this way in far future since external electromagnetic signals, not included in the 
initial data on $S$, will interfere. As is well known, the field is uniquely 
determined by the data on $S$ only in the spacetime region which on a two--dimensional 
diagram (see Fig. 1) is represented by a ,,triangle'' whose base is $S$ and the other 
two sides are future directed null lines (photon paths) emitted from the boundary 
points of $S$. This region is termed the \textit{domain of dependence in future\/} of 
$S$, $D^+(S)$, or the \textit{future Cauchy development\/} of $S$. In Minkowski 
spacetime the hypersurface $S$ may be extended to the entire physical space (in an 
inertial frame) $t=t_0$, then the electromagnetic field (and other physical fields) 
is uniquely determined for arbitrarily distant future (and past), i.~e.~for all times. 
This is possible because the conformal infinity consists there of two null cones and 
no external signal can enter the spacetime from outside (i.~e.~from $\mathcal{J}$) 
without crossing the space at $t=t_0$. Also in many curved spacetimes there exist 
spacelike hypersurfaces (being sets of simultaneous events with respect to some 
coordinate time) which, if treated as initial data surfaces, allow to predict to 
whole future and past. 

\begin{figure}[ht!]
 \includegraphics[scale=0.8]{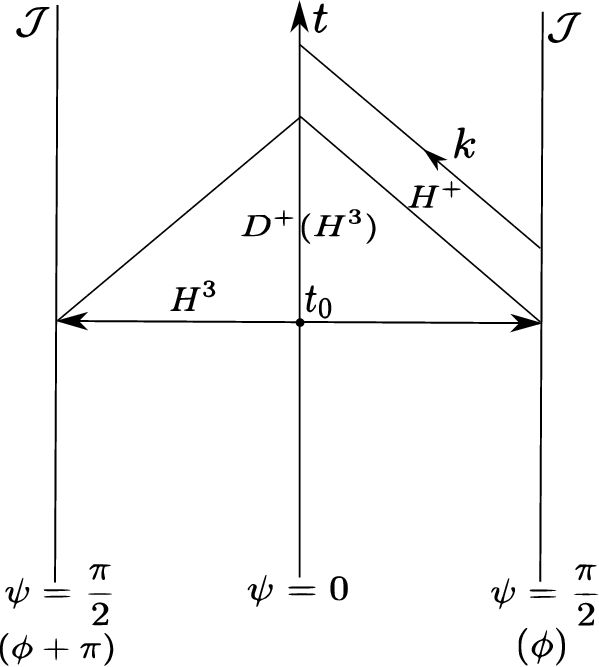}
  \caption{Two--dimensional representation of CAdS space for $\theta=\pi/2$. 
  Besides points with $\psi=0$ where there is a coordinate singularity, each point 
  $(t,\psi)$ represents a half circle in $\phi$. The boundary cylinder $\mathcal{J}$ 
	is depicted as one line for some $\phi$ and as the antipodal line at $\phi+\pi$. 
	The initial data hypersurface is the whole space $H^3$ at $t=t_0$. The null 
	boundaries $H^+$ of $D^+(H^3)$ are null future lines emanating from the boundary 
	infinity $S^2$ of $H^3$. Any electromagnetic signal $k$ entering CAdS from 
	$\mathcal{J}$ for $t>t_0$ moves outside $D^+(H^3)$ and affects the solution along 
	its path.}
\end{figure}											
This is not the case of CAdS space. In Fig. 1 the diagram in coordinates $(t-\psi)$ is 
presented. $\theta=\pi/2$ and each point of the diagram to the left and right of the line 
$\psi=0$ represents a half--circle of coordinate $\phi$. The line $\psi=0$ consists of 
single points because the coordinate system is singular there and the spheres of 
$(\theta,\phi)$ shrink to a point. The boundary $\mathcal{J}$ is shown as two lines, 
one for some fixed value of $\phi$ and 
the other as opposite to it, $\phi+\pi$. One takes the whole space $H^3$ for some $t_0$ as 
the initial data surface, then any physical field is uniquely determined in the domain of 
dependence $D^+(H^3)$ bounded in the future by two null hypersurfaces $H^+$ made of null 
rays emanating from the sphere being the intersection of $H^3$ with $\mathcal{J}$. The 
region $D^+(H^3)$ cannot cover the whole CAdS because any light signal emitted from 
$\mathcal{J}$ at $t>t_0$ will perturb the field. Physics in CAdS is unpredictable. This is 
particularly troublesome for quantizing fields propagating in this world \cite{AIS}. 

\section{Uniformly accelerating observers}
The static coordinate system of eqs. (4), (5) and (7) may be given a physical 
interpretation by showing that observers at rest, $r,\theta,\phi=\textrm{const}$,  
are actually uniformly accelerating ones \cite{Po}. The notion of uniform 
acceleration is taken directly from special relativity (SR). In SR consider a motion 
of a particle in a fixed inertial frame of reference denoted by LAB. In this frame the 
particle has 4--velocity $u^{\alpha}=dx^{\alpha}/ds=(\gamma, \gamma \mathbf{v}/c)$ 
and 4--acceleration 
\begin{equation}\label{n17}
w^{\alpha}\equiv\frac{du^{\alpha}}{ds}=
\frac{1}{c^2}\gamma^4\left[\frac{1}{c}\mathbf{v}\cdot
\mathbf{a}, (\frac{1}{c}\mathbf{v}\cdot\mathbf{a})\frac{\mathbf{v}}{c}+
\frac{1}{\gamma^2}\mathbf{a}\right],
\end{equation} 
where the Lorentz factor is $\gamma=(1-\mathbf{v}^2/c^2)^{-1/2}$ and $\mathbf{a}=
d\mathbf{v}/dt$ is the ordinary 3--acceleration measured in LAB. The identity 
$u^{\alpha}u_{\alpha}=\eta_{\alpha\beta}u^{\alpha}u^{\beta}=1$, where 
$\eta_{\alpha\beta}=\textrm{diag}[+1,-1,-1,-1]$ is the Minkowski metric, implies 
$\eta_{\alpha\beta}u^{\alpha}w^{\beta}=0$ and $w^{\alpha}$ is a spacelike vector with 
the squared length
\begin{equation}\label{n18}
\eta_{\alpha\beta}w^{\alpha}w^{\beta}=-\frac{\gamma^4}{c^4}\left[\gamma^2
(\frac{1}{c}\mathbf{v}\cdot\mathbf{a})^2+\mathbf{a}^2\right]<0.
\end{equation}  
Whereas LAB is an arbitrary frame, the particle has a distinguished inertial frame, 
the local proper frame in which it is momentarily at rest. In the proper frame the 
particle has $\mathbf{v}=\mathbf{0}$ and the acceleration is denoted by $\mathbf{a}= 
\mathbf{A}$; in consequence $w^{\alpha}w_{\alpha}=-\mathbf{A}^2/c^4$. The particle 
is \textit{uniformly accelerated\/} if $\mathbf{A}=\vec{const}$ and in the case 
of a one--dimensional motion it amounts to $w^{\alpha}w_{\alpha}=\textrm{const}<0$. 
In any curved spacetime again $u^{\alpha}=dx^{\alpha}/ds$ and the acceleration 
vector is the absolute derivative with respect to $s$ of the velocity vector,
\begin{equation}\label{n19}
w^{\alpha}=\frac{D}{ds}\frac{dx^{\alpha}}{ds}\equiv \frac{d^2x^{\alpha}}{ds^2}+
\Gamma^{\alpha}_{\mu\nu}\frac{dx^{\mu}}{ds}\frac{dx^{\nu}}{ds},
\end{equation} 
where $\Gamma^{\alpha}_{\mu\nu}$ are the Christoffel symbols for the metric 
$g_{\alpha\beta}(x^{\mu})$. Again $g_{\alpha\beta}u^{\alpha}u^{\beta}=1$ and 
$g_{\alpha\beta}u^{\alpha}w^{\beta}=0$. Take the CAdS metric as in eq. (7) and a 
static observer with $\psi=\psi_0>0$ and $\theta,\phi=\textrm{const}$. Then 
along its worldline $ds=dt/\cos\psi_0$ and $u^{\alpha}=(dx^{\alpha}/dt)(dt/ds)=
[\cos\psi_0,0,0,0]$. In the static coordinates the observer remains at rest and is 
uniformly accelerated iff $w^{\alpha}w_{\alpha}=\textrm{const}<0$. One needs not 
to compute the Christoffel symbols since the covariant components of the 
acceleration are given by 
\begin{equation}\label{n20}
w_{\alpha}=\frac{D}{ds}u_{\alpha}=\frac{d}{ds}(g_{\alpha\beta}u^{\beta})-
\frac{1}{2}g_{\mu\nu,\alpha}u^{\mu}u^{\nu}.
\end{equation} 
One gets $w_{\alpha}=-\delta^1_{\alpha}\tan\psi_0$ and $w^{\alpha}=+\frac{1}{a^2}
\delta^{\alpha}_1\sin\psi_0\cos\psi_0$, then $w^{\alpha}w_{\alpha}=-\frac{1}{a^2}
\sin^2\psi_0$ and identifying this expression with $-A^2$ (one returns to $c=1$) 
one finds that each static observer is subject to a uniform acceleration equal to 
$\frac{1}{a}\sin\psi_0$. The acceleration monotonically grows with $\psi_0$ and 
reaches maximum at the spatial infinity. For $\psi_0=0$ the acceleration vanishes. 

\section{Geodesic lines}
In a curved spacetime the geodesic lines play the same role as straight lines do in 
euclidean spaces. The straight line has two fundamental properties: i) the vector 
tangent to it at any point, when parallelly transported along it to any other point, 
remains tangent to it, and ii) it is the shortest line between any pair of its points. 
The second feature cannot be implemented without some changes in a spacetime. Already 
in Minkowski spacetime a straight timelike line is the \textit{longest\/} one between 
its points. The timelike geodesic \textit{maximizes\/} the spacetime interval between 
its points. Along the null geodesic, as along any other null curve, the interval 
between any pair of points, is zero. Only the spacelike geodesic is the shortest line 
joining two points. Yet the first property is transferred unaltered into any 
spacetime: the geodesic is such a line that for any parametric representation of the 
line, $x^{\alpha}=x^{\alpha}(v)$, $\alpha=0,1,2,3$, the acceleration vector (i.~e.~the 
absolute derivative with respect to $v$ of the tangent vector) is proportional to the 
tangent vector,        
\begin{equation}\label{n21}
\frac{D}{dv}\frac{dx^{\alpha}}{dv}\equiv\frac{d^2x^{\alpha}}{dv^2}+\Gamma^{\alpha}_
{\mu\nu}\,\frac{dx^{\mu}}{dv}\frac{dx^{\nu}}{dv}=h(v)\,\frac{dx^{\alpha}}{dv},
\end{equation} 
where $h(v)$ is a scalar function depending on the choice of 
the parameter $v$. The proportionality feature is exactly as in mechanics: a body in 
a rectilinear motion may either move uniformly, if the temporal parameter $t$ is 
appropriately chosen, or move non-uniformly with respect to a different parameter 
$t'$, say $t=\ln t'$. Guided by this analogy one can show that there exists such a 
parametrization of the geodesic that the acceleration vanishes, $h(v)\equiv 0$, then 
$v$ is termed \textit{canonical parameter\/}. For a timelike geodesic the canonical 
parameter coincides with the arc length (the proper time), $v=s$; for spacelike 
geodesics the parameter denoted by $l$ is defined as $dl^2=-ds^2>0$ and for null 
ones the parameter $\sigma$ has no simple geometrical or physical interpretation. 
In practice one replaces eq. (21) (for the canonical $v$) by the equivalent form 
which avoids computing $\Gamma^{\alpha}_{\mu\nu}$ symbols and arises from eq. (20),
\begin{equation}\label{n22}
\frac{d}{dv}\left(g_{\alpha\beta}\,\frac{dx^{\beta}}{dv}\right)-\frac{1}{2}\,
g_{\mu\nu,\alpha}\,\frac{dx^{\mu}}{dv}\frac{dx^{\nu}}{dv}=0.
\end{equation}
The behavior of the three types of geodesics exhibits the fundamental geometrical 
properties of the spacetime under consideration.\\

We begin studying geodesics in CAdS space with the spacelike ones. We use them to 
determine the distance from any given point to the spatial infinity $\mathcal{J}$. 
The distance is defined as the length of a spacelike geodesic joining the given 
point $P_0$ at $t=t_0$ to any simultaneous point at $\mathcal{J}$. We use the 
reference frame in which the metric is explicitly static, eqs. (4), (5) or (7), 
hence we expect that all points of the geodesic are simultaneous, $t=t_0$. Since 
CAdS is spherically symmetric, we expect that the geodesic is radial, i.~e.~the 
angles $\theta$ and $\phi$ are constant along it and only the radial coordinate is 
variable. One then need not at all to solve the geodesic equation, it suffices to 
compute the length of the radial line. Using e.~g.~eq. (4) one gets the distance 
from $r_0$ to $r_1$ equal $l(r_0,r_1)=r_1-r_0$. 
The distance from any internal point to $\mathcal{J}$ is infinite, as it should be 
expected.\\
From the explicit form of the geodesic equation one infers that circular spacelike 
geodesics, $r=\textrm{const}>0$ and $\theta=\pi/2$, do not exist.

\section{Null geodesics}
Interpreting any null geodesic as a worldline of a photon (being in the classical 
approximation a point particle) and the tangent vector as the wave vector, one writes 
$x^{\alpha}=x^{\alpha}(\sigma)$ and $dx^{\alpha}/d\sigma=k^{\alpha}$, then the 
geodesic equation reads 
\begin{equation}\label{n23}
\frac{d}{d\sigma}\left(g_{\alpha\beta}\,k^{\beta}\right)-\frac{1}{2}\,
g_{\mu\nu,\alpha}\,k^{\mu}k^{\nu}=0.
\end{equation}
The canonical parameter is determined up to a linear transformation (change of 
units), hence one may assume that $\sigma$ is dimensionless. One knows from section 
2 that CAdS space is conformally flat. In general, if two spacetimes are conformally 
related, then they have the same null geodesics (in the sense of the same null lines). 
In fact, assume that $\bar{g}_{\mu\nu}=\Omega^2\,g_{\mu\nu}$ and eq. (23) holds. 
Then making an appropriate transformation of the canonical parameter, $\bar{\sigma}
=f(\sigma)$ with $f'>0$, one shows by a direct calculation that the transformed 
wave vector $\bar{k}^{\alpha}=dx^{\alpha}/d\bar{\sigma}$ satisfies the same equation 
for the rescaled metric,
\begin{equation}\label{n24}
\frac{d}{d\bar{\sigma}}\left(\bar{g}_{\alpha\beta}\,\bar{k}^{\beta}\right)-\frac{1}{2}\,
\bar{g}_{\mu\nu,\alpha}\,\bar{k}^{\mu}\bar{k}^{\nu}=0.
\end{equation}
The function $f(\sigma)$ is determined by $\Omega$ via a differential equation. 
Applying the Poincar\'{e} coordinates, eq. (11) and (12), one sees that null 
geodesics of CAdS coincide with those of Minkowski space in coordinates $x^{\alpha}=
(t',x,y,z)$; these are straight lines $x^{\alpha}=a^{\alpha}\bar{\sigma}+
\textrm{const}$, where a constant vector $a^{\alpha}=(a^0,\mathbf{a})$ is null, 
$(a^0)^2-\mathbf{a}^2=0$, and $-\infty<\bar{\sigma}<\infty$. Instead of determining 
$\sigma=f^{-1}(\bar{\sigma})$ we directly solve the geodesic equation in the global 
coordinate system, eq. (4). We consider a radial geodesic $x^{\alpha}=(t(\sigma),
r(\sigma), \pi/2,0)$, then $k^{\alpha}=(dt/d\sigma, dr/d\sigma,0,0)$. Since 
$(\partial/\partial t)g_{\mu\nu}=0$, eq. (23) for $\alpha=0$ is immediately 
integrated, 
\begin{equation}\label{n25}
\frac{d}{d\sigma}\left(k^0\cosh^2\frac{r}{a}\right)=0 \Rightarrow \frac{dt}
{d\sigma}\,\cosh^2\frac{r}{a}=\textrm{const}\equiv Ea>0,
\end{equation}
where a dimensionless $E$ is proportional to the conserved energy of the photon. 
The equations for $\theta$ and $\phi$ hold identically and the second order 
equation for $r$ is replaced by the constraint\\
$g_{\alpha\beta}\,k^{\alpha}k^{\beta}=0=\dot{t}^2\cosh^2r/a-\dot{r}^2$\\
working as an integral of motion 
and assuming that the geodesic emanates from $r=r_0\geq 0$ for $\sigma=0$ with 
$\dot{r}=dr/d\sigma>0$, and employing eq. (25) one gets
\begin{equation}\label{n26}
E\sigma=\sinh\frac{r}{a}-\sinh\frac{r_0}{a}, \qquad 
r=a\ln\left(E\sigma+\sinh\frac{r_0}{a}+\sqrt{(E\sigma+\sinh\frac{r_0}{a})^2+
1}\,\right).
\end{equation}
It is convenient to use also 
the angular radial variable of eq. (7), $\tan\psi=\sinh r/a$, then one finds 
$E\sigma=\tan\psi-A$, where $A\equiv\sinh r_0/a=\tan\psi_0$.  Due to the conformal 
invariance of the null geodesic equation (23), the solution is independent of 
the cosmological constant $\Lambda=-3/a^2$. 
A radial photon emanating from any point reaches the spatial infinity $\mathcal{J}$ 
for $\sigma\rightarrow\infty$, as expected. This means that $\mathcal{J}$ consists 
of endpoints of future and past directed radial null geodesics and coincides with 
the set of endpoints of radial spacelike geodesics. Yet integrating eq. (25) one 
gets $t(\sigma)$ and the simplest expression arises if the variables $\psi$ and 
$\psi_0$ are used,                         
\begin{equation}\label{n27}
\frac{dt}{d\sigma}=Ea[(E\sigma+A)^2+1]^{-1} \quad \textrm{and} \quad  
t-t_0=a\arctan(E\sigma+A)-a\psi_0,
\end{equation}
or $t(\sigma)-t_0=a(\psi-\psi_0)$. The light cone in the variables $(t,a\psi)$ 
consists of straight lines inclined at $45^{\circ}$, as in Minkowski space. 
The coordinate time interval of the photon flight from $\psi=\psi_0$ to $\mathcal{J}$ 
is \textit{finite\/} and its maximum value is $t-t_0=\pi a/2$ for $\psi_0=0$.   
Let the photon be emitted from point $A$, $t=t_A$ and $\psi=\psi_0$, moves 
radially outwards, reaches the spatial infinity where it is reflected by a mirror 
and returns to $\psi=\psi_0$ at the event $B$ at $t=t_B$, Fig. 2. The time of 
the flight is finite, $t_B-t_A=(\pi-2\psi_0)a$, though the 
distance from $\psi_0$ to $\mathcal{J}$ (measured along a spacelike radial 
geodesic) is infinite, $l(\psi_0,\pi/2)=\infty$. 
Also the proper time $s$ measured by a clock staying at $\psi=\psi_0$ between the 
emission and return of the photon is finite; from eq. (7) one has 
\begin{equation}\label{n28}
ds^2=\frac{dt^2}{\cos^2\psi_0} \quad \Rightarrow\quad s(A,B)=\frac{t_B-t_A}
{\cos\psi_0}=\frac{\pi-2\psi_0}{\cos\psi_0}\,a.
\end{equation}
$s(A,B)$ decreases from $\pi a$ for $\psi_0=0$ to $2a$ for $\psi_0\rightarrow \pi/2$.

\begin{figure}[ht!]
\includegraphics[scale=0.8]{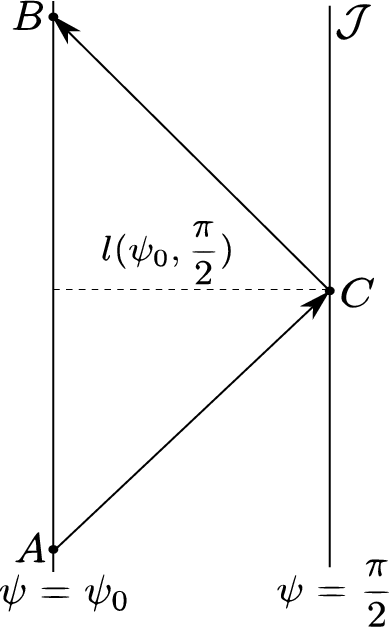}
  \caption{The photon is emitted from A, $t=t_A$ and $\psi=\psi_0$, radially 
	outwards, reaches the spatial infinity at C, where it is reflected by a mirror 
	and returns to $\psi=\psi_0$ at B, $t=t_B$. The time of flight, $t_B-t_A$, and 
	the proper time $s(A,B)$ are both finite and bounded from above, though the distance 
	to $\mathcal{J}$ is $l(\psi_0,\pi/2)=\infty$.}
\end{figure}	
 
What kind of a curve in the ambient space $\mathbf{R}^{3,2}$ is the radial null 
geodesic of eqs. (26) and (27)? By rotations of the spheres one can always put 
$\theta=\pi/2$ and $\phi=0$ along the geodesic, then applying eqs. (3) one finds 
the parametric description of the geodesic in the ambient space (we employ the 
relationships between functions arctan, arcsin and arccos),
\begin{equation}\label{n29}
U=Ea\sigma, \quad V=a, \quad X=Ea\sigma, \quad Y=Z=0.
\end{equation}
This is a straight line which is null, since the tangent five--vector\\ 
$(dU/d\sigma,\ldots,dZ/d\sigma)=Ea(1,0,1,0,0)$ is null in the metric of eq. (1). 
It is well known that in euclidean 3--space the one--sheeted hyperboloid $x^2+
y^2-z^2=1$ contains a 1--parameter family of straight lines, which are geodesic 
curves on both the hyperboloid and in the space. Analogously, AdS space contains a 
1--parameter family of null geodesics (the parameter is the energy $E$) being 
null straight lines of the ambient $\mathbf{R}^{3,2}$.\\

In Schwarzschild spacetime generated by a static star or a static black hole there 
exists one (unstable) circular null geodesic: if a photon is emitted from a point 
on the equator of the sphere with the radial coordinate $r_0=3GM/c^2$ in a direction 
tangent to the equator, the gravitational field of the central body of mass $M$ will 
capture it and the photon will revolve for ever around it on the circular orbit. In 
CAdS space one verifies, using the metric of eq. (4), that circular null geodesics 
do not exist for any finite value of the radial variable $r$. In fact, the radial 
component of the geodesic equation, i.~e.~the $\alpha=1$ component of equation (23), 
together with the integral of motion $g_{\alpha\beta}k^{\alpha}k^{\beta}=0$ show 
that the assumption $r=\textrm{const}=r_0$ is consistent only if $\sinh r_0=
\cosh r_0$, or $r_0=\infty$. Formally, a circular null geodesic exists only at the 
spatial infinity.\\ 
Properties of null geodesics are to some extent related to the problem of stability 
of CAdS space. Spacetimes that approach CAdS one at infinity are called 
\textit{asymptotically CAdS spacetimes\/} (a rigorous definition is quite 
sophisticated). It has been shown that CAdS space is a ground state for asymptotically 
CAdS spacetimes, in the same sense as Minkowski space is the ground state for 
spacetimes which are asymptotically flat. In any field theory the ground state 
solution must be stable against small perturbations, otherwise the theory is 
unphysical. For Minkowski space it has been proven after long and sophisticated 
investigations that the space is stable since sufficiently small initial 
perturbations vanish in distant future due to radiating off their energy to 
infinity. The spatial infinity $\mathcal{J}$ of CAdS space actually is a timelike 
hypersurface and any radiation may either enter the space through $\mathcal{J}$ 
or escape through it. It is therefore crucial for the question of stability to 
correctly choose a boundary condition at infinity. Most researchers assume 
\textit{reflective boundary conditions\/}: there is no energy flux across the conformal 
boundary $\mathcal{J}$, in other terms the boundary acts like a mirror at which 
outgoing fields (perturbations) bounce off and return to the interior of the 
spacetime. Under this assumption P. Bizo\'{n} recently received a renowned result: 
CAdS space is unstable against formation of a black hole for a large class of 
arbitrarily small perturbations \cite{Bi}.\\
We have a critical remark to this outcome. The instability is due to the presence of 
matter in the form of the linear massless scalar field and it is physically relevant 
provided it is not a peculiarity specific to the scalar field. The instability must also 
develop for dust matter and electromagnetic perturbations (this has not been checked 
yet due to computational difficulties). Suppose that the instability is triggered by 
high frequency electromagnetic waves of small amplitude, these may be viewed as 
photons. Consider a photon belonging to the perturbation. As is depicted in Fig. 2 
the outgoing photon is subject at point C to the reflective boundary conditions and 
is forced to come back. Since CAdS space is maximally symmetric, the photon has 
conserved both its energy and linear momentum. For the incoming (returning) photon 
the spatial momentum has the opposite sign to that of the outgoing photon and this 
is possible only if the photon meets at C a physical mirror and is bounced off it. 
In other terms the reflective boundary conditions mean that CAdS space is equivalent 
to a box with material walls. Gravitational instability of perturbations closed in 
a box is less surprising.

\section{Timelike geodesics}
Consider a cloud of free test particles, each of unit rest mass, whose own 
gravitational field is negligible, which move in CAdS space. The notion of 
,,negligible'' is intuitively clear, but in the framework of GR it is based on a 
deeper reasoning. First, in GR a point particle does not exist: a point particle 
with a mass, no matter how small, actually is a black hole with the event horizon 
and diverging curvature near the singularity. Therefore a ,,point particle'' is an 
approximation and means an extended body of a diameter $d$ and one assumes that all 
distances under consideration have scale $L\gg d$. In this sense GR is similar to 
celestial mechanics where planets are viewed as pointlike objects provided the error 
$L$ of determining their orbits is much larger than their diameters. If $L\approx d$ 
one must take into account the physical nature of the object. Second, one compares 
the gravitational field (the curvature) of the particle of mass $m$, computed at 
the distance $L$ from it, to the external gravitational field, in the present case 
being the CAdS space curvature. If the external curvature is much larger than that of 
each particle, their gravitation is negligible and the particles are viewed as ,,test'' 
ones. (In consequence, in the flat spacetime, particles are free and test ones only 
if their gravitational interactions are completely neglected.) Assuming that this is 
the case, each particle moves on a timelike geodesic of the CAdS metric. 
Let one choose a reference frame adapted to the cloud: the 
frame is comoving with the particles, what means that every particle has constant 
spatial coordinates, then its worldline coincides with one coordinate time line. 
Though the particles are ,,motionless'' in this frame, the distances between them 
vary in time as the cloud expands or shrinks, hence in the frame the metric is time 
dependent. In this comoving frame the CAdS metric has the form given in eq. (14). 
In fact, let a particle of the cloud be at rest in the coordinate system $(\tau, 
\chi, \theta,\phi)$. Then along its worldline there is $ds^2=d\tau^2$ or $\tau-
\tau_0=s$ and the tangent vector is $u^{\alpha}=dx^{\alpha}/ds=dx^{\alpha}/d\tau=
(1,0,0,0)$. The timelike geodesic equation, according to eqs. (21) and (22), is 
\begin{equation}\label{n30}
\frac{d}{ds}\left(g_{\alpha\beta}\,u^{\beta}\right)-\frac{1}{2}\,
g_{\mu\nu,\alpha}\,u^{\mu}u^{\nu}=0
\end{equation}
and for this worldline it holds identically since it reduces to \\
$\frac{d}{ds}\,g_{\alpha 0}-\frac{1}{2}\,g_{00,\alpha}\equiv 0$.\\
The curves $\tau-\tau_0=s$ and $\chi,\theta,\phi=\textrm{const}$ are timelike 
geodesics and as such these are the worldlines of free particles (actually the 
coordinate time lines in the comoving system are geodesic worldlines also in the case of a 
self--gravitating cloud of particles, but then the metric differs from that of CAdS 
space). One notices that these geodesics are orthogonal to the physical spaces $H^3$ 
given by $\tau=\textrm{const}$; this is why this comoving frame is termed 
\textit{Gaussian normal geodesic\/} (GNG) system. Furthermore the time coordinate 
$\tau$ is the physical time measured by good clocks travelling along these geodesics 
since it is equal to intervals of proper time, $\Delta \tau=\Delta s$.\\

In a generic spherically symmetric spacetime one usually singles out the simplest 
timelike geodesic curves, radial and circular. These are geometrically distinguished 
by the symmetry centre and their distinction is frame independent. (In de Sitter 
space the circular geodesics do exist, but they are revealed in the GNG coordinates, 
whereas the frequently used static coordinates, covering only a half of the manifold, 
deceptively suggest that circular geodesics are excluded) \cite{SG1}. It is therefore 
rather astonishing that in CAdS space the difference between radial and circular 
geodesics is merely coordinate dependent and geometrically they form the same curve. 
Furthermore, each ,,generic'' timelike geodesic may be transformed into a circular 
or a radial one. The proof of that using ,,internal'' methods, that is the 
four--dimensional metric of the space, is complicated and we shall apply the external 
approach based on the use of the ambient flat space $\mathbf{R}^{3,2}$.\\
To this end we resort to AdS space as a pseudosphere in $\mathbf{R}^{3,2}$ since this 
piece of CAdS is sufficient (recall that CAdS is the infinite chain of AdS spaces 
opened in the time direction and glued together). One describes any timelike geodesic 
G on AdS space as a curve in the embedding $\mathbf{R}^{3,2}$. Using the coordinates 
$X^A=(U,V,X,Y,Z)$, $A=1,\ldots 5$, the curve G is parametrized by its length, $X^A=
X^A(s)$. Clearly G is not a geodesic (a straight line) of the flat ambient space 
$\mathbf{R}^{3,2}$. The geodesic equation follows from a variational principle and its 
derivation may be performed in the ambient space, the result reads (see e.~g.~ \cite
{SG1})
\begin{equation}\label{n31}
\ddot{X}^A+\frac{1}{a^2}X^A=0,
\end{equation} 
where $\dot{X}^A=dX^A/ds$. These are five decoupled equations and their general 
solution depends on ten arbitrary constants. The solution describes the geodesic G if 
it satisfies two constraints, the definition of AdS space given in eq. (2) and the 
normalization of the velocity five--vector $\dot{X}^A$, $\dot{U}^2+\dot{V}^2-
\dot{X}^2-\dot{Y}^2-\dot{Z}^2=1$. The geodesic is then
\begin{equation}\label{n32}
X^A(s)=q^A\sin(\frac{s}{a}+c)+p^A\cos(\frac{s}{a}+c),
\end{equation} 
where $c$ is an integration constant and two constant directional five--vectors 
$q^A$ and $p^A$ (constancy of the two vectors has a geometrical meaning because the 
ambient space is flat and the coordinates $X^A$ are Cartesian) are subject to three conditions,
\begin{equation}\label{n33}
q^Aq_A=a^2, \qquad p^Ap_A=a^2 \quad \textrm{and} \quad q^Ap_A=0,
\end{equation} 
here $q^Ap_A=\eta_{AB}q^Ap^B$ and $\eta_{AB}=\textrm{diag}[1,1,-1,-1,-1]$ is the metric 
tensor in eq. (1). Altogether the arbitrary geodesic G, eq. (32), depends on eight 
initial values. \\
Now one puts $c=0$ for simplicity and employs the full SO(3,2) symmetry of 
$\mathbf{R}^{3,2}$. Let $P_0\in\mathbf{R}^{3,2}$ be an initial point ($s=0$) of G. 
Take any transformation of SO(3,2) which makes the coordinates of $P_0$ equal to 
$X=Y=Z=U=0$ and $V=a$, the transformation is non--unique. Then by the remaining 
transformations leaving invariant the straight line joining $P_0$ with the origin 
$X^A=0$ one makes the tangent to G at $P_0$ vector $\dot{X}^A(0)$ tangent to the $U$ 
line through $P_0$, i.~e.~ $\dot{U}(0)=1$ and $\dot{V}(0)=\dot{X}(0)=\dot{Y}(0)=
\dot{Z}(0)=0$. Then the representation of G is reduced to 
\begin{equation}\label{34}
U(s)=a\sin\frac{s}{a}, \qquad V(s)=a\cos\frac{s}{a}, \qquad X=Y=Z=0.
\end{equation}
Each timelike geodesic on AdS space is represented in $\mathbf{R}^{3,2}$ by a circle 
of the same radius $a$ (determined by the curvature of the space) on an appropriately 
chosen euclidean two--plane $(U,V)$ \cite{SG1}. The distinction between radial, 
circular and ,,general'' geodesics has no geometrical meaning and in this space 
there is only one kind of timelike geodesics, analogously to Minkowski space 
possessing only one geodesic, a timelike straight line, which may be identified with 
the time axis of an inertial reference frame. (Recall that Minkowski space arises 
in the limit $a\rightarrow\infty$.) In other terms each timelike geodesic of AdS 
space is the circle lying on a euclidean two--plane going through the origin $X^A=0$ 
of the ambient space. In general two timelike geodesics do not intersect and this 
means that their two--planes do not intersect either and the planes have only one 
common point, the origin.\\
One can also find an explicit transformation in $\mathbf{R}^{3,2}$ recasting a 
circular geodesic into a radial one, see Appendix. We emphasize that these 
properties of timelike geodesics are easy to investigate in the embedding flat 
five--space, whereas the internal four--dimensional approach is rather difficult. 

\section{Further properties of timelike geodesics}
First we draw an important conclusion from the fact that each geodesic on AdS is 
the circle in $\mathbf{R}^{3,2}$. Accordingly, the parametric description in eq. 
(32) shows that each geodesic is periodic with the period $\Delta s=2\pi a$ 
corresponding to one turn around the circle. We now analytically show that any two 
timelike geodesics emanating from an arbitrary point of AdS space first diverge 
and then reconverge at the distance $s=\pi a$, again diverge from that point and 
finally return to the initial point in the ambient space for $s=2\pi a$. Let two 
arbitrary geodesics, $G_1$ and $G_2$, emanate from an arbitrary point $P_0$. One 
chooses the coordinates $X^A$ adapted to $P_0$ and $G_1$: the coordinates of $P_0$ 
are $X^A(P_0)=(a,0,0,0,0)$ and the directional vectors of $G_1$ are directed along 
the axes $X^1=U$ and $X^2=V$ respectively, $p_1^A=(a,0,0,0,0)$ and $q_1^A=(0,a,0,0,0)$. 
Then $G_1$ is 
\begin{equation}\label{n35}
X_1^A(s)=q_1^A\sin\frac{s}{a}+p_1^A\cos\frac{s}{a}.
\end{equation} 
This implies that $G_2$ has a generic form of eq. (32) with the vectors $q^A$ and 
$p^A$ related by 
\begin{equation}\label{n36}
p^1=\frac{1}{\cos c}(a-q^1\sin c), \qquad p^i=-q^i \tan c, \qquad i=2,3,4,5.
\end{equation} 
The three conditions in eq. (33) imply that $(q^1)^2$ is determined by $q^i$ and 
$\sin c=q^1/a$, thus arbitrary $G_2$ starting from $P_0$ is determined by four 
arbitrary parameters $q^i$, corresponding to four independent components of the 
initial velocity $\dot{X}^A(0)$. One sees from eqs. (32) and (35) that at the 
distance $s=\pi a$ counted along both the geodesics one has $X_1^A(\pi a)=-p_1^A$ 
for $G_1$ and from eq. (36) one has $X^A(\pi a)=-(a,0,0,0,0)=X_1^A(\pi a)$ for 
$G_2$, or the two geodesics intersect at this point. This is a \textit{point 
conjugate\/} to $P_0$ on $G_1$ and $G_2$ and antipodal to $P_0$ in $\mathbf{R}^{3,2}$. 
At the distance $s=2\pi a$ both the geodesics return to $P_0$, $X_1^A(2\pi a)=
X^A(2\pi a)= (a,0,0,0,0)$, or make a closed loop on the pseudosphere in 
$\mathbf{R}^{3,2}$.\\
Geometrically this effect is obvious. $G_1$ and $G_2$ are circles of the same 
radius lying on two--planes $\pi_1$ and $\pi_2$ respectively. Since $P_0$ is the 
common point of the circles, $\pi_1$ and $\pi_2$ intersect along the straight line 
connecting $P_0$ to the origin. Then the antipodal to $P_0$ point $P_1$ (i.~e.~
having $X^A(P_1)=-X^A(P_0)$) lies on this line and $G_1$ and $G_2$ go through $P_1$ 
after delineating a half--circle from $P_0$.\\ 

In CAdS space the periodic time is replaced by the infinite line. For timelike 
geodesics this implies that each geodesic does not return to the initial point at 
the distance $\Delta s=2\pi a$, but goes to a new point, which is the same point 
in the three--space (using the static coordinates of eqs. (4), (5) and (7)) and 
is shifted forward in the time. The geodesics have in CAdS space infinite extension, 
yet their relationships cannot be altered in comparison to these in AdS space. 
Two geodesics having a common initial 
point must intersect for $\Delta s=2\pi a$ and the intersections will repeat 
infinitely many times, always after the same interval of the proper time. It 
turns out that it is hard to show this effect in full generality using exclusively 
the internal four--dimensional description due to computational difficulties. 
Geometrically it is clear that it is sufficient to show the effect for radial 
geodesics (the case including circular geodesics is discussed in the next section) 
and to this end the comoving coordinates $(\tau, \chi, \theta, \phi)$ of eq. (14) 
are most appropriate. Consider the geodesics orthogonal to $\tau=\textrm{const}$ 
hypersurfaces, these are the coordinate time $\tau$ lines, $\chi,\theta,\phi=
\textrm{const}$. The distance between two neighboring geodesics (simultaneous 
points) is 
\begin{equation}\label{n37}
dl^2= a^2\sin^2\frac{\tau}{a}\,(d\chi^2+\sinh^2\chi\,d\Omega^2)
\end{equation}
and is largest for $\tau=\pi a/2$ and tends to zero both in the past for $\tau
\rightarrow 0$ and in future for $\tau\rightarrow\pi a$. This means that all these 
hypersurface orthogonal geodesics emanate from the common point $\tau=0$ and diverge 
until $\tau=\pi a/2$, then reconverge at $\tau=\pi a$. The comoving coordinates are 
valid in the region between two hypersurfaces, $\tau=0$ and $\tau=\pi a$, which 
metrically shrink to one point. In CAdS space these coordinates hold independently 
in each region between $\tau=n\pi a$ and $\tau=(n+1)\pi a$ for any integer $n$; 
together these regions form an infinite chain, which, as we saw in Sect. 2, cover 
only a small part of the entire manifold. The fact that the geodesics actually 
intersect after $\Delta \tau=\pi a$ rather than $2\pi a$ corresponds to the 
geometrical effect in AdS space that the circles intersect twice. The points 
$\tau=n\pi a$ form an infinite sequence of points conjugate to $\tau=0$ along 
these geodesics.\\
We emphasize that although CAdS space is static with timelike lines infinitely 
extending and it is a solution to Einstein field equations which may be constructed 
without the intermediating stage of the pseudosphere in the flat five--space, 
nevertheless this geodesic reconvergence is a residual effect of the time periodicity 
of the AdS space as the pseudosphere. Without invoking the pseudoshere in 
$\mathbf{R}^{3,2}$ this property of CAdS space is incomprehensible.\\

The fact that in CAdS space all timelike geodesics starting from a common point can 
only recede from each other to a finite distance and then must intersect infinite many 
times, has two important consequences. First, a timelike geodesic cannot reach the 
spatial infinity $\mathcal{J}$. In fact, the infinity is for $r$ and $\rho\rightarrow
\infty$ and according to eq. (3) all the coordinates $X^A$ are infinite there (except 
for discrete values of $t$, $\theta$ and $\phi$ where some $X^A$ vanish). Yet it is 
seen from eq. (32) that the coordinates $X^A(s)$ of a timelike geodesic are always 
finite. Another, purely four--dimensional proof in the case of a radial geodesic is 
given in sect. 9.\\
Second, there are points inside the future light cone of any $P_0$ that cannot be 
reached from $P_0$ by any timelike geodesic. We are accustomed to in Minkowski space 
and expect the same effect in any curved spacetime (as it occurs in the Schwarzschild 
field) that if two points can be connected 
by a timelike curve, they can also be connected by a geodesic. This is not the case of 
CAdS space. To show it we again employ the five--dimensional description since one AdS 
space is sufficient to this aim. Let a bunch of geodesics emanate from arbitrary $P_0$. 
We have seen that at the distance $\Delta\tau=\pi a=\Delta s$ from $P_0$ all geodesics 
intersect at $P_1$. Take any spacelike 3--dimensional hypersurface S through $P_1$. 
The future light cone from $P_0$ intersects S along a closed surface $\Sigma$ (having 
topology of the two--sphere) being a boundary of a 3--dimensional set D in S. The set 
D, lying in the interior of the light cone, belongs to the \textit{chronological 
future\/} of $P_0$, i.~e.~any point of D may be connected to $P_0$ by a timelike curve. 
However, no point of D besides $P_1$, can be connected to $P_0$ by a geodesic. In 
other words, a large part (an open region) of the interior of the future (past) light 
cone of $P_0$ is inaccessible from $P_0$ along a timelike geodesic.

\section{The twin paradox}
Finally we discuss a version of the twin paradox known from special relativity (SR). 
In SR the ,,paradox'' has a purely geometrical nature and consists in determining the 
longest timelike curve joining two given points P and Q (providing Q lies in the 
chronological future of P). There is no shortest curve since a timelike curve from 
P to Q may have arbitrarily small length. The solution in SR is simple: it is the 
straight line connecting P to Q. Physically this means that the twin which gets older 
at the reunion is the twin which always stays   
at rest in the inertial reference frame where this line is a coordinate time line. 
In a curved spacetime the problem is more sophisticated since there are actually 
two separate problems: a local and a global one. In CAdS space, due to its maximal 
symmetry, the two problems coincide. We consider three twins (,,siblings''): twin A 
stays at rest at a fixed point in space, twin B revolves on a circular geodesic 
orbit around a chosen origin of spherically symmetric coordinates and twin C moves 
upwards and downwards on a radial geodesic in these coordinates. Their worldlines 
emanate from a common initial point $P_0$ and we study where they will intersect in 
the future \cite{SG2}. We apply the static coordinates $(t,\rho,\theta,\phi)$ of 
eq. (5).\\
The nongeodesic twin A remains at $\rho=\rho_0>0$, $\theta=\pi/2$ and $\phi=0$ and 
in a coordinate time interval $T$ its worldline has length
\begin{equation}\label{n38}
s_A(T)= \sqrt{\left(\frac{\rho_0}{a}\right)^2+1}\,T.
\end{equation} 
Any twin following a geodesic has conserved energy and denoting its energy per unit 
mass by $k$ (dimensionless) one finds \cite{SG2}
\begin{equation}\label{n39}
\dot{t}\equiv \frac{dt}{ds}=\frac{a^2k}{\rho^2+a^2}.
\end{equation}
As the initial point we choose $P_0(t_0=0,\rho=\rho_0>0, \theta=\pi/2,\phi=0)$. For 
the circular geodesic of B with $\rho=\rho_0$ one has $t=s$, $\phi=s/a=t/a$ and its 
energy is related to the radius by $\rho_0=\sqrt {k_B-1}a$. The period of one 
revolution is $T=2\pi a$ and the length of B for one revolution is $s_B=T=2\pi a$, 
the already known result. After one revolution the twins A and B meet and $s_A(2\pi a)
>s_B$, or there are timelike curves longer than the geodesic B.\\
The twin C moving on a radial geodesic has the radial velocity $\dot{\rho}\equiv 
d\rho/ds$ given by 
\begin{equation}\label{n40}
\dot{\rho}^2=k_C^2-\left(\frac{\rho^2}{a^2}+1\right)
\end{equation}
following from $g_{\alpha\beta}u^{\alpha}u^{\beta}=1$. Let at $P_0$ twin C be initially 
at rest, $\dot{\rho}(0)=0$, then its energy is $k_C^2=(\rho_0/a)^2+1$ and from the 
radial component of the geodesic equation (30) it follows that its acceleration is 
directed downwards, $\ddot{\rho}(0)=-(\rho_0/a)^2<0$ and the twin falls down. This 
shows that gravitation in CAdS space is attractive. (This is not trivial since in de 
Sitter space gravitational forces are repulsive.) We therefore consider a more general 
motion: C radially flies away with $\dot{\rho}(0)=u>0$, reaches a maximum height 
$\rho=\rho_M$, falls down back to $\rho=\rho_0$ and then to $\rho=0$ and farther 
(for $\phi=\pi$). The highest point of the trajectory is, from eq. (40), $\rho_M^2=
(k_C^2-1)a^2$, and $\rho_M>\rho_0$ implies $k_C^2>(\rho_0/a)^2+1$. One sees that a 
radial geodesic cannot reach the spatial infinity $\mathcal{J}$ since $\rho_M\rightarrow
\infty$ requires infinite energy $k_C\rightarrow\infty$. Moving in the opposite 
direction ($\phi=\pi$), C reaches the same highest point, $\rho=\rho_M$, and falls 
down back to $(\rho_0,\phi=0)$; in this way it oscillates between the antipodal in the 
3--space points $(\rho_M, \phi=0)$ and $(\rho_M, \phi=\pi)$ infinite many times. The 
coordinates of the geodesic C may be parametrically described by $t=t(\eta)$, $\rho=
\rho(\eta)$ and $s=s(\eta)$, where $\eta$ is an angular parameter \cite{SG2}, here 
we use a simpler description. To this end we again apply the five--dimensional picture. 
The points of the geodesic have coordinates $X^A$ given in eq. (3), where one puts 
$\rho=a\sinh r/a$ and $\theta=\pi/2$ and $\phi=0$, then 
\begin{equation}\label{n41}
U=\sqrt{\rho^2(s)+a^2}\sin\frac{t}{a}, \quad V=\sqrt{\rho^2(s)+a^2}\cos\frac{t}{a},
\quad X=\rho(s), \quad Y=Z=0.
\end{equation}
On the other hand C is described by eq. (32) with $c=0$. By comparing the two expressions 
one finds $X=\rho=q^1\sin s/a+p^1\cos s/a$. To determine $q^1$ and $p^1$ one inserts 
this expression into both eq. (40) and into the radial component of the geodesic equation 
(30) and checks that it is a solution to these equations. Applying the initial conditions 
one gets 
\begin{equation}\label{n42}
\rho(s)=\sqrt{\rho_M^2-\rho_0^2}\,\sin\frac{s}{a}+ \rho_0\cos\frac{s}{a}
\end{equation}
and the highest point $\rho_M$ is reached for $\cos s_M/a=\rho_0/\rho_M$. Since the 
domain of the radial coordinate is $\rho\geq 0$, the values $\rho(s)<0$ are assigned to 
points with $\rho=|\rho(s)|$ and $\phi=\pi$. Clearly, the proper time 
interval between the highest points, $(\rho_M,\phi=0)$ and $(\rho_M,\pi)$ is $\Delta s=
\pi a$ and the same interval is between the initial point $(\rho_0,0)$ and its 
antipodal one $(\rho_0,\pi)$, independently of the energy $k_C$ \cite{SG2}.  
Notice that the special solution $\rho(s)=0$ and $k=1$ actually represents the 
,,canonical'' description of any timelike geodesic given in eq. (34). This shows 
that the detailed behavior of any geodesic revealed by the general solution in eq. (42) 
is merely coordinate dependent.\\
One can also integrate eq. (39) applying eq. (42) but then one gets a generic formula 
for $t(s)$ being a complicated expression involving functions arc tan, which is of little 
use. Instead one considers a special case of $\rho_0=0$, then eq. (42) is reduced to 
\begin{equation}\label{n43}
\rho(s)=\sqrt{k^2-1}\,a\sin\frac{s}{a}
\end{equation}
and if the length of this geodesic is divided into intervals according to 
$s=(\sigma+\frac{1}{2}n\pi)a$, where $0\leq 
\sigma<\pi/2$ and $n=0,1,2,\ldots$, the time coordinate is
\begin{equation}\label{n44}
t(s)=a\arctan(k\tan\sigma)+\frac{1}{2}n\pi a.
\end{equation} 
The coordinate time interval between the highest point, $\sqrt{k^2-1}a$, and its 
antipodal one (always in the 3--space), is $\Delta t=\pi a$, and clearly the same 
holds for the generic geodesic C. This means that the geodesics B and C will intersect 
first at $(t=\pi a, \phi=0)$ and then at $(t=2\pi a, \phi=0)$ and later infinite many 
times. Whereas the twins B and C meet each other after the constant intervals 
$\Delta s=\pi a=\Delta t$, which are independent of C's initial velocity, twin C 
meets A after the time interval \cite{SG2}
\begin{equation}\label{n45}
\Delta t_1=\pi a-2a\arctan\left(\frac{k_C\rho_0}{\sqrt{\rho_M^2-\rho_0^2}}\right)
\end{equation} 
and the corresponding length of the geodesic C is
\begin{equation}\label{n46}
s_C=2a\arccos\left(\frac{\rho_0}{\rho_M}\right).
\end{equation}  
These two expressions are so complicated that it is not easy to analytically compare 
the lengths of A and C for the interval $\Delta t_1$ (being the time of C's flight 
on the route $\rho_0\rightarrow\rho_M\rightarrow\rho_0$). It has been numerically 
shown that always $s_C>s_A(\Delta t_1)$, as it should be, since on this segment of 
the geodesic C there are no points conjugate to $P_0$. In Fig. 3 we depict two 
radial geodesics emanating from a common point.

\begin{figure}[ht!]
\includegraphics[scale=0.8]{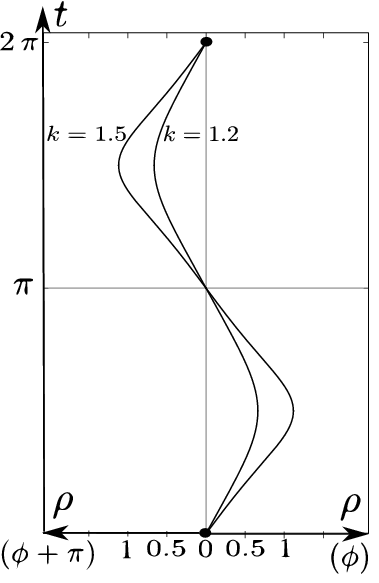}
  \caption{Two radial timelike geodesics with energies $k=1.2$ and $k=1.5$, 
	emanating from a common point, chosen for simplicity as $t=0=\rho$. The length 
	scale is $a=1$. The geodesics must return to $\rho=0$ and intersect at $t=\pi$, 
	then they go to $\rho<0$ corresponding to the direction $\phi+\pi$ and again 
	intersect at $t=2\pi$, $\rho=0$. This evolution repeats infinite many times. 
	All points $(t=\pi,\rho\neq 0)$ and $(t=2\pi,\rho\neq 0)$ are inaccessible from 
	the initial point along a timelike geodesic.}
\end{figure}	

Finally we illustrate the results of the paper with a numerical example. Let the 
curvature scale of CAdS space be $a=10^{16}\textrm{m}=1$ light year, then the 
cosmological constant is $\Lambda=-3/a^2=-3\cdot 10^{-32}\textrm{m}^{-2}$. Let from 
the coordinate origin $\rho_0=0$ be emitted simultaneously a photon and a particle 
of mass $m$, both radially and in the same direction, they move along a null and a 
timelike geodesic, respectively. At a spatial point P at $\rho=\rho_M=10^6 a$ they are 
reflected backwards by a mirror and return to $\rho_0=0$. Employing formulae given above 
one finds that along the photon path there is $\rho(\sigma)=Ea\sigma$ and the value 
$\rho_M$ at the highest point shows that $E\sigma=10^6$, then the distance between 
the origin and P (measured along the radial spacelike geodesic at $t=\textrm{const}$) 
is $l(0,\rho_M)\cong a\ln(2\rho_M/a)\cong 14,5a=14,5$ l.y. The coordinate time of the 
photon flight to P and back is equal to the proper time measured by a clock staying 
at rest at $\rho_0=0$ and is $\Delta t=\Delta s=2a\arctan(E\sigma)$ and is very slightly 
below $\pi a$ or slightly below $\pi$ years. The massive particle is ultrarelativistic 
and closely follows the photon; from $\rho_M=\sqrt{k^2-1}\,a$ one gets that its energy 
is $10^6mc^2$ and its proper time interval when reaching $\rho_M$ is exactly 
$s_M=\pi a/2$, then its total travel lasts $2s_M=\pi$ years. Both the photon and the 
particle travel the distance $2\cdot 14,5=29$ l.y. and to go it they need a period of 
time not exceeding $\pi$ years. This outcome deceptively suggests that the photon 
and the particle move at superluminal velocities since their average velocity is 
$29/\pi\cong 9,2c$. Clearly the local velocity of light is always $c$ and this 
superluminal one is merely a result of the weird geometry of CAdS space.
 
\section{Conclusions}
Anti--de Sitter space is one of the three simplest, maximally symmetric solutions to 
vacuum Einstein field equations. Its metric is static with the time coordinate 
extending from $-\infty$ to $+\infty$, nonetheless most of its geometric properties 
are periodic in the time, something which is incomprehensible from the intrinsic 
four--dimensional viewpoint. The light seems to move at superluminal velocities since 
the photon may travel over arbitrarily large distances (to spatial infinity and back) 
in a finite time interval. In static coordinates covering the whole spacetime one can 
single out in the set of all timelike geodesics the radial and circular curves, yet it 
turns out that this distinction is geometrically irrelevant and is merely coordinate 
dependent. No timelike geodesic can escape to the spatial infinity unless it has 
infinite energy. Also a timelike geodesic may travel large distances at a 
superluminal average velocity. All timelike geodesics emanating from a common 
initial event $(t_0,\mathbf{x})$ return to the same point $\mathbf{x}$ in the space 
after the time interval $\Delta t=2\pi a$; this means that all simultaneous events 
($t=t_0+2\pi a$), though belonging to the chronological future of the initial event, 
are inaccessible from the latter by a timelike geodesic. In other words, any point 
$(t_0+2\pi a,\mathbf{y})$ cannot be reached from $(t_0,\mathbf{x})$ by a free fall 
in any direction and with any initial velocity, if the points $\mathbf{x}$ and 
$\mathbf{y}$ are different. These bizarre features become understandable only if 
one divides the whole spacetime into an infinite chain of segments and each of them 
is identified with the anti--de Sitter space proper and the latter is modelled as a 
pseudosphere in an unphysical five--dimensional space. In this space each timelike 
geodesic od AdS space forms a circle of the same radius, which accounts for their 
weird properties. This necessity is in conflict with general relativity stating 
that a physical spacetime is four--dimensional and all its properties are 
intrinsically grounded, without resorting to a fictitious higher dimensional 
embedding space. Finally, if the boundary conditions are suitably chosen, AdS 
space is unstable and cannot be a ground state for spacetimes with $\Lambda<0$. 
The conclusion, therefore, is unambiguous: this spacetime is unphysical and cannot 
describe a physical world. It may only serve as a mathematical tool in field 
theory, e.~g.~in the recent AdS/CFT correspondence.\\

\textbf{Acknowledgments}. We gratefully acknowledge useful critical comments by 
Andrzej Staruszkiewicz and we are grateful to Szymon Sikora for help in preparing 
the figures.

\section{Appendix}
Here we derive an explicit transformation recasting a circular timelike geodesic 
on AdS space into a radial one. 
To this end one uses the static coordinates $(t,\rho,\theta,\phi)$ of eq. 
(5), then points $X^A$ of the pseudosphere in the ambient space are parametrized by 
these variables according to eq. (3) with $\rho=a\sinh r/a$. For any timelike geodesic 
the angles $\theta$ and $\phi$ may be so chosen that the curve lies in the 
two-surface $\theta=\pi/2$, then its points are 
\begin{equation}\label{n47}
U=\sqrt{\rho^2+a^2}\sin\frac{t}{a}, \quad V=\sqrt{\rho^2+a^2}\cos\frac{t}{a}, \quad 
X=\rho\cos\phi, \, Y=\rho\sin\phi, \, Z=0.
\end{equation}
As in sect. 9 a circular geodesic $G_c$ has $\rho=\rho_0>0$, $t=s$, $\phi=s/a=t/a$, 
its radius is determined by the energy, $\rho_0=\sqrt{k_c-1}\,a$ and its coordinates 
$X_c^A$ are  
\begin{equation}\label{n48}
X_c^A(s)=q_c^A\sin\frac{s}{a}+p_c^A\cos\frac{s}{a}.
\end{equation} 
To determine the directional five--vectors one compares eq. (47) for $\rho=\rho_0$, 
$t=s$ and $\phi=s/a$ with eq. (48) and gets $p_c^A=(0,\sqrt{k_c}\,a,\sqrt{k_c-1}\,a,0,0)$ 
and $q_c^A=(\sqrt{k_c}\,a,0,0,\sqrt{k_c-1}\,a,0)$.\\
Now assume that in a Cartesian coordinate system $X'^A$ (different from $X^A$ one) 
a radial geodesic $G_r$ is described by 
\begin{equation}\label{n49}
X_r^{'A}(s)=q_r^A\sin\frac{s}{a}+p_r^A\cos\frac{s}{a}
\end{equation}
and $X_r^{'A}$ are parametrized by $x'^{\alpha}=(t',\rho',\theta',\phi')$ as in eq. (47). 
One sets $\phi'(s)=0$ at points of $G_r$ and assuming that it emanates from $\rho'(0)=
0$ with $\dot{\rho}'(0)>0$ its coordinates $\rho'(s)$ and $t'(s)$ are given by the 
right--hand sides of eqs. (43) and (44), the latter holds for $n=0,1,2,3$. To 
determine the directional vectors in this case it is sufficient to take $n=0$ in eq. 
(44) and apply the identity $\arctan x\equiv \arcsin[x(1+x^2)^{-1/2}]$, then 
\begin{equation}\label{n50}
\sin\frac{t'}{a}=\frac{k\sin\frac{s}{a}}{(\cos^2\frac{s}{a}+k^2\sin^2\frac{s}{a})^{1/2}}.
\end{equation}
Next one inserts the relationships (47) into eq. (49) with due replacements of $X^A$ 
by $X'^A$, $x^{\alpha}$ by $x'^{\alpha}$ and with $\phi'=0$ and employs there eq. (43) 
for $\rho'$ and eq. (50). Finally the normalizations of eq. (33) provide 
$q_r^A=(ak_r,0,\sqrt{k_r^2-1}\,a,0,0)$ and $p_r^A=(0,a,0,0,0)$.
If $G_r$ and $G_c$ are two different (coordinate dependent) 
descriptions of the same curve in $\mathbf{R}^{3,2}$, there exists a linear 
transformation of the pair $(q_c^A,p_c^A)$ into $(q_r^A,p_r^A)$. One then seeks for a 
transformation $L\in SO(3,2)$, $X'^A=L^A{}_BX^B$ such that $L^A{}_B q_c^B=q_r^A$ 
and $L^A{}_B p_c^B=p_r^A$. According to the fundamental theorem both the geodesics 
are geometrically represented by circles with the same radius, hence all other their 
characteristics, such as the conserved energy, are coordinate dependent and irrelevant. 
One can therefore put $k_c=k_r\equiv k$. A simple and long computation results in 
$L$ depending on one arbitrary parameter and setting it equal zero one gets the 
simplest form of the matrix,
\begin{equation}\label{n51}
(L^A{}_B)=\left(\begin{array}{ccccc}
k^{3/2}         & (k-1)\sqrt{k+1} & -\sqrt{k(k^2-1)}  & -k\sqrt{k-1}     & 0 \\
  0             & \sqrt{k}        &     -\sqrt{k-1}   &   0              & 0 \\
\sqrt{k(k^2-1)} & k\sqrt{k-1}     &  -k^{3/2}         & -(k-1)\sqrt{k+1} & 0 \\
   \sqrt{k-1}   &  0              &      0            &  -\sqrt{k}       & 0 \\
   0            &     0           &      0            &       0          & 1   
\end{array}\right),     
\end{equation}
$\det L=+1$. 
It is clear that both $G_c$ and $G_r$ emanate from the same point for $s=0$. In fact, 
the initial point $P_0$ of $G_c$ has coordinates $X^A(P_0)=X_c^A(0)=p_c^A$ and after the 
transformation its coordinates are $X'^A=L^A{}_B p_c^B=p_r^A$ and these are the 
coordinates of the initial point of $G_r$, $X_r^{'A}(0)=p_r^A$. We notice that the  
transformation in AdS space from $(t,\rho,\phi)$ to $(t',\rho',\phi')$ (for $\theta=
\theta'=\pi/2$) is very intricate and hence useless.


\begin{thebibliography}{99}
\bibitem{Hu}
V. E. Hubeny, ``The AdS/CFT correspondence'', Class. Quantum Grav. \textbf{32} 
(2015) 124010 (42 pp) [arXiv:1501.00007 [gr-qc]].
\bibitem{CM} 
E. Calabi and L. Markus, ``Relativistic space forms'', Annals of 
Mathematics \textbf{75} (1962) 63--76.
\bibitem{GP}
J. Griffiths and J. Podolsk\'{y}, \textit{Exact space--times in Einstein's 
general relativity} (Cambridge Univ. Press, Cambridge, 2009), chap. 5.
\bibitem{HE}
S. W. Hawking and G.F. R. Ellis, \textit{The large scale structure of space--time} 
(Cambridge Univ. Press, Cambridge, 1973).
\bibitem{Pe}
R. Penrose, \textit{Structure of space--time. Battelle Rencontres, 1967 Lectures in 
Mathematics and Physics} (W. Benjamin, New York, 1968).
\bibitem{AIS}
A. Avis, C. Isham and D. Storey, ``Quantum field theory in anti--de Sitter 
space--time'', Phys. Rev. \textbf{D18} 3565--3576.
\bibitem{Po}
J. Podolsk\'{y}, ``Accelerating black holes in anti--de Sitter universe'', Czech. 
J. Phys. \textbf{52} (2002) 1--10 [arXiv:gr-qc/0202033].
\bibitem{Bi}
P. Bizo\'{n}, ``Is AdS stable?'', Gen. Relativ. Gravit. (2014) 46: 1724.
\bibitem{SG1}
L. M. Soko\l{}owski and Z. A. Golda, ``Every timelike geodesic in anti--de Sitter 
spacetime is a circle of the same radius'', Intern. J. Mod. Phys. \textbf{D25} 
(2016) 1650007 (6 pp). 
\bibitem{SG2}
L. M. Soko\l{}owski and Z. A. Golda, ``The local and global geometrical aspects of the 
twin paradox in static spacetimes: I. Three spherically symmetric spacetimes'', 
Acta Phys. Polon. \textbf{B45} (2014) 1051--1075 [arXiv:1402.6511v2 [gr-qc]].

\end{thebibliography}
\end{document}